\def\authorversion{}
\documentclass[conference]{IEEEtran}

\usepackage[utf8]{inputenc}
\usepackage{cite}
\usepackage{amsmath}
\usepackage{amsfonts}
\usepackage{amssymb}
\usepackage{graphicx}
\usepackage{csquotes}
\usepackage{tikz}
\usepackage{tikz-3dplot}
\usepackage{hyperref}
\usepackage{booktabs}
\usepackage{pgfplots}
\usepackage{rwthstyle}
\usepackage{array}

\usepackage{iksmath}

\usepackage{./tikzstyle}
\usetikzlibrary{positioning}
\usetikzlibrary{iksblockdiagrams}
\usetikzlibrary{calc}

\ifdefined\authorversion
	\usepackage{lastpage}
	\usepackage{eso-pic}
	\AddToShipoutPicture{
		\put(0,25){%
			\hspace{1in}\hspace{\oddsidemargin}
			\begin{minipage}{\paperwidth-2in-\oddsidemargin-\oddsidemargin}
				\scriptsize
				International Conference on Immersive and 3D Audio (I3DA), Sept. 2021, Bologna, Italy (online).\\
				\textsc{DOI}: \href{https://doi.org/10.1109/I3DA48870.2021.9610968}{10.1109/I3DA48870.2021.9610968}.
				Pre-print available at \url{https://arxiv.org/abs/2104.13069}.
				\\[1ex]\copyright{} 2021 IEEE. %
				\hfill \thepage\,/\,\pageref*{LastPage}
			\end{minipage}%
		}
	}
\fi

\graphicspath{{figures/}}

\makeindex

\newlength\figureheight
\newlength\figurewidth
\newcommand{\figuredatadir}{figures-data/}

\newcommand{\trans}{^\mathrm{T}} %
\newcommand{\herm}{^\mathrm{H}}  %
\newcommand{\conj}{^*}           %
\newcommand{\inv}{^{-1}}         %

\DeclareMathOperator{\trace}{tr}

\newcommand{\az}{\phi}    %
\newcommand{\inc}{\theta} %

\newcommand{\uvec}[1]{\vec{#1}_\text{u}} %
\newcommand{\unitsphere}{{\mathcal{S}^2}}

\newcommand{\sigin}{x}
\newcommand{\sigout}{\tilde{x}}
\newcommand{\sigexc}{u}
\newcommand{\sigresp}{\tilde{u}}

\newcommand{\shcoeff}[1]{{#1}_{nm}}
\newcommand{\shcoeffvec}[1]{\boldsymbol{{#1}_{nm}}}

\newcommand{\puvec}[1]{\vec{#1}_u\mkern-6mu\vphantom{#1}}

\newcommand{\crazylinelegend}{\tikz[x=2em]{\draw[ultra thick] (0,0) -- (1,0); \draw[yellow,thick] (0,0) -- (1,0); \fill[blue] (0,0) circle(0.5ex); \fill[red] (1,0) circle(0.5ex);}}

\title{Visualization of Linear Operations in the Spherical Harmonics Domain}

\author{%
	\IEEEauthorblockN{%
		Maximilian Kentgens,
		Peter Jax
	}
	\\
	\IEEEauthorblockA{%
		\textit{RWTH Aachen University, Germany} \\
		\{kentgens, jax\}@iks.rwth-aachen.de
	}
}

\begin{document}

	\makeatletter
	\hypersetup{pdftitle={\@title},pdfauthor={Maximilian Kentgens, Peter Jax}, hidelinks}
	\makeatother
	
	\maketitle
	
	\begin{abstract}
		Linear operations on coefficients in the spherical harmonics (SH) transform domain that again yield SH-domain coefficients are an important toolset in many disciplines of research and engineering. They comprise rotations, spatially selective filters, and many other modifications for various applications, or describe the response of a MIMO system to an excitation.
		It is of particular importance to characterize these operations both qualitatively and quantitatively, and make them accessible for people to work with.
		In this paper, we identify different key properties of such operations and propose a method for their visualization. With our proposed method, we succeed to show many important aspects of an operation in a single plot and give rise to a comprehensive interpretation of the behavior of a system.
		In our evaluation, we show the potential of the proposed method on the basis of various practical examples from spatial audio signal processing, where SH-domain filtering is used to modify acoustic scenes given by higher-order Ambisonics signals.
	\end{abstract}

	\begin{IEEEkeywords}
		spherical harmonics, visualization, signal processing on the sphere, higher-order Ambisonics, linear operations, spatial filtering
	\end{IEEEkeywords}
	
	\section{Introduction}

	Spherical harmonics (SH) arise in various disciplines of science and engineering whenever a wave equation is solved by separation of variables in spherical coordinates. They find interdisciplinary applications including acoustics, electromagnetism, quantum mechanics, computer graphics, as well as geodesy and earth science. %
	SH series offer an elegant way to represent square-integrable functions defined on the unit sphere $\unitsphere$ as the SH functions form a complete orthonormal basis~\cite{GumerovDuraiswami2004,ArfkenWeber2005}. In practice, only the band-limited expansion up to a certain truncation order $N$ is considered resulting in a finite number of $(N+1)^2$ coefficients. The latter can be stacked in a vector $\shcoeffvec{\sigin}$ for notational convenience.
	
	Without loss of generality, we here focus on the field of audio signal processing, where the SH domain is frequently used to encode the angular pattern of sound wave incidence into a reference point in space~\cite{Williams1999}.
	This approach is used both in the beamforming community in research on spherical sensor arrays~\cite{Rafaely2015} and in the spatial audio community who coined the term \textit{higher-order Ambisonics} for these kinds of signals~\cite{Daniel2001}.

\begin{figure}
		\centering
		\input{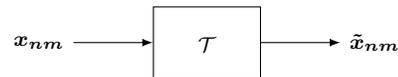}%
\needsPackage{amsmath}%
\needsPackage{rwthstyle}
\needsTikzlibrary{iksblockdiagrams}%
\needsTikzlibrary{positioning}%
\needsTikzlibrary{calc}%
\providecommand{\unitsize}{3em}%
\begin{tikzpicture}[
	node distance=\unitsize,
	font=\footnotesize,
	>=latex,
	block/.style={iks block,minimum height=6ex,minimum width=9ex},
]
	\node (src) {$\boldsymbol{x_{nm}}$};
	\node[block,right=of src] (transf) {$\mathcal{T}$};
	\node[right=of transf] (snk) {$\boldsymbol{\tilde{x}_{nm}}$};
	
	\draw [->] (src)   -- (transf);
	\draw [->] (transf) -- (snk);	
\end{tikzpicture}%
\tikzaloneEnd%
		\caption{Block diagram}
		\label{fig:block_diagram}
	\end{figure}

	In practice, a single set of coefficients occurs rarely but a series of coefficient vectors over an additional dimension such as time or frequency is often considered. Nevertheless, many spatial operations are performed on each coefficient vector individually. In spatial audio processing, for example, this is due to the fact that in the respective transformation domain, time and space can be considered orthogonal, and the operations are therefore memoryless. As a consequence, we neglect any additional dependency of the coefficient vector. 
	An arbitrary linear operation $\mathcal{T}$ defined by a transformation matrix $\boldsymbol{T}$ can be simply expressed as
	\begin{align}
		\label{eq:sh_transformation}
		\shcoeffvec{\sigout} = \boldsymbol{T} \shcoeffvec{\sigin}
	\end{align}
	where $\shcoeffvec{\sigout}$ is again a set of SH coefficients like the input $\shcoeffvec{\sigin}$. This is also depicted in Fig.~\ref{fig:block_diagram}.
	Not only simple operations such as rotation and mirroring~\cite{Wigner1931,GumerovDuraiswami2004,ArfkenWeber2005}, a directional loudness modification~\cite{KronlacherZotter2014}, or other static filters~\cite{ZotterPomberger2011,KentgensJax2020} but also a variety of complex time-variant adaptive filter methods and aggregate systems such as a SH noise reduction~\cite{HerzogHabets2019} or a perceptually motivated modification operation~\cite{KentgensBehlerJax2020} fit into this versatile framework~\cite{KronlacherZotter2014,HafsatiEpainDaniel2017}.
	
	Classic filter theory for one-dimensional signals has a long history of tools for the visualization of the properties of a filter. Magnitude, phase and group-delay plots, and sometimes just the structure of the time-domain filter coefficients give deep insights in the behavior of a system.
	Although most people would not deny that these standard tools are of tremendous value in research and education, there is to the best of our knowledge no such equivalent for a comprehensive visualization of linear transformations in the SH domain. Instead, researchers tend to visualize the response for a typical system excitation or develop very much specialized methods for their specific problem.
	
	The objective of the present work is to propose a unified method to visualize the most important aspects of any linear SH operation in a single plot aiming at a maximum of both comprehensiveness and clarity. The necessary fundamentals including a discussion of how such operations can be characterized are provided in Sec.~\ref{sec:fundamentals}. The proposed method follows in Sec.~\ref{sec:proposed_method}. The potential and possible limitations of the method are evaluated and discussed on the basis of real-world examples in Secs.~\ref{sec:evaluation} and \ref{sec:discussion}, respectively.

	\section{Linear Operations in the Spherical Harmonics Domain}
	\label{sec:fundamentals}
	
	In this section, we recap the SH fundamentals and discuss how any operation of the form of Eq.~\eqref{eq:sh_transformation} can be characterized by its responses to a set of excitation signals. The latter will be the base for the proposed visualization method.
	
	\subsection{Spherical Harmonics}
	
	\begin{figure}
		\centering
		\tikzset{font=\footnotesize}
		\def\scale{1.7}
		\input{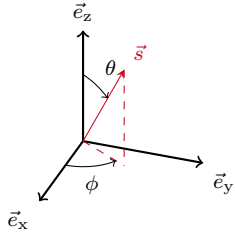}
\needsPackage{tikz-3dplot}%
\needsPackage{rwthstyle}%
\providecommand{\scale}{3}%
\tdplotsetmaincoords{60}{110}%
\pgfmathsetmacro{\rvec}{1.0}%
\pgfmathsetmacro{\thetavec}{30}%
\pgfmathsetmacro{\phivec}{60}%
\begin{tikzpicture}[scale=\scale,tdplot_main_coords]
	\coordinate (O) at (0,0,0);
	\draw[thick,->] (0,0,0) -- (1,0,0) node[anchor=north east]{$\vec{e}_\mathrm{x}$};
	\draw[thick,->] (0,0,0) -- (0,1,0) node[anchor=north west]{$\vec{e}_\mathrm{y}$};
	\draw[thick,->] (0,0,0) -- (0,0,1) node[anchor=south]{$\vec{e}_\mathrm{z}$};
	\tdplotsetcoord{P}{\rvec}{\thetavec}{\phivec}
	\draw[-stealth,color=rwthrot] (O) -- (P) node[above right] {$\vec{s}$};
	\draw[dashed, color=rwthrot] (O) -- (Pxy);
	\draw[dashed, color=rwthrot] (P) -- (Pxy);
	\tdplotdrawarc[->]{(O)}{0.4}{0}{\phivec}{anchor=north}{$\phi$}
	\tdplotsetthetaplanecoords{\phivec}
	\tdplotdrawarc[tdplot_rotated_coords,->]{(0,0,0)}{0.6}{0}{\thetavec}{anchor=south west}{$\theta$}
\end{tikzpicture}%
\tikzaloneEnd
		\vspace{-1.5ex}
		\caption{Definition of the spherical coordinate system}
		\label{fig:coordinate_system}
	\end{figure}
	
	The SH coefficient vector $\shcoeffvec{\sigin}$ as introduced with \eqref{eq:sh_transformation} consists of the $(N+1)^2$ coefficients $\shcoeff{\sigin}$ with order $n=0,1,\ldots,N$ and degree $m$ with $-n \le m \le n$. They can be transformed to the spatial domain using spherical harmonics $Y_n^m(\uvec{s})$ as basis functions,~\cite{Rafaely2015}
	\begin{align}
		\label{eq:idsht}
		\sigin(\inc,\az) \equiv
		\sigin(\uvec{s})
		&=  \sum_{n=0}^N \sum_{m=-n}^n \sigin_{nm} Y_n^m(\uvec{s}).
	\end{align}
	The notation $\uvec{\scalebox{0.8}[0.7]{(\!}\cdot\scalebox{0.8}[0.7]{\!)}}$ denotes a normalization of a Cartesian vector to unit length. Accordingly, $\uvec{s}\in\mathbb{R}^3$ is the Cartesian equivalent of a spherical coordinate $(\inc,\az)$ defined on the unit sphere $\unitsphere$. The definition of the inclination angle $\inc \in [0,\pi]$ and azimuth angle $\az \in (-\pi,\pi]$ is shown in Fig.~\ref{fig:coordinate_system}. Both variants are used for convenience in the following.
	
	The inverse operation of \eqref{eq:idsht} is
	\begin{align}
		\shcoeff{\sigin} = \int_{\unitsphere} \sigin(\uvec{s}) [Y_n^m(\uvec{s})]\conj \diff{}\uvec{s},
	\end{align}
	where $\diff{}\uvec{s} \equiv \sin\inc \diff\inc \diff\az$ is the surface element on the unit sphere~$\unitsphere$ and $(\cdot)\conj$ denotes the complex conjugate.

	Evaluating $\sigin(\uvec{s})$ on a grid $\mathcal{Q} \subset \unitsphere$ with $Q$ sampling points allows to formulate a discrete version of \eqref{eq:idsht} in matrix notation, i.e.,
	\begin{align}
		\label{eq:idsht_vec}
		\boldsymbol{\sigin} = \boldsymbol{Y}_{N,\mathcal{Q}} \shcoeffvec{\sigin}.
	\end{align}
	The SH matrix is defined as $\boldsymbol{Y}_{N,\mathcal{Q}} = \left[ \boldsymbol{y}_{N}\trans(\vec{s}_{\text{u},1}), \ldots, \boldsymbol{y}_{N}\trans(\vec{s}_{\text{u},Q})  \right]\trans$, where each SH vector $\boldsymbol{y}_{N}(\vec{s}_{\text{u},q})$ is a row vector for the $q$-th direction in $\mathcal{Q}$. The notation $(\cdot)\trans$ denotes the matrix transpose.

	\subsection{Characterization of Linear Operations}
	\label{sec:linear_operation_characterization}
	
	We assume that Eq.~\eqref{eq:sh_transformation} is a transformation from a signal of truncation order $N$ to $\tilde{N}$. Thus, the dimension of $\boldsymbol{T}$ is $(\tilde{N}+1)^2 \times (N+1)^2$.
	
	The operation $\mathcal{T}$ can be considered a \textit{multiple-input and multiple-output} (MIMO) system. To identify such a system, one requires at least $(N+1)^2$ orthogonal excitation signals for which the responses yield a complete system description.
	Since the SH are a set of orthogonal functions, it is obvious to use them as an excitation signal. When sampling the SH on $\unitsphere$, spatial aliasing has to be avoided~\cite{RafaelyWeissBachmat2007}. The SH matrix $\boldsymbol{Y}_{N,\mathcal{Q}}$ is orthogonal for a \mbox{(quasi-)}uniform \mbox{t-design} sampling grid $\mathcal{Q}$ with a sufficiently high number of $Q$ sampling points such that  $\frac{4\pi}{Q}\boldsymbol{Y}_{N,\mathcal{Q}}\herm \boldsymbol{Y}_{N,\mathcal{Q}} = \boldsymbol{I}$ holds. Here, $\boldsymbol{I}$ is the identity matrix and $(\cdot)\herm$ denotes the matrix Hermetian. In general, $Q>(\tilde{N}+1)^2$ but the exact minimum value of $Q$ strongly depends on the respective choice of sampling grid. For details regarding the sampling theorems on the sphere, the reader shall be referred to e.g.,~\cite{DriscollHealy1994,Rafaely2015}.
	
\begin{figure}
		\centering
		\setlength{\abovecaptionskip}{0pt} %
		\setlength{\belowcaptionskip}{-10pt} %
		\tikzset{every axis/.append style={grid=both}}
		\setlength\figurewidth{\linewidth}
		\setlength\figureheight{.48\linewidth}
		\input{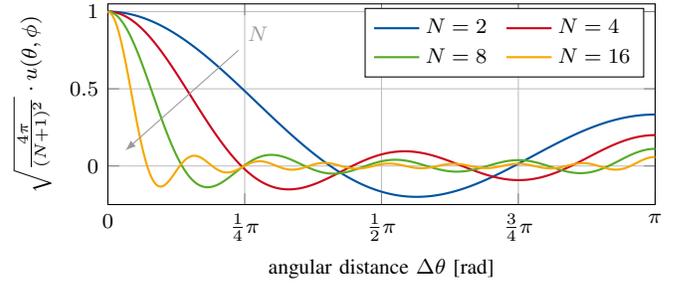}%
\needsPackage{../tikzstyle}%
\providecommand{\figurewidth}{\axisdefaultwidth}%
\providecommand{\figureheight}{\axisdefaultwidth}%
\providecommand{\figuredatadir}{../figures-data/}%
\begin{tikzpicture}
[
	font=\footnotesize,
	>=latex
]
	\begin{axis}
	[
		width=\figurewidth,
		height=\figureheight,
		xlabel={angular distance $\Delta\theta$ [rad]},
		ylabel={$\sqrt{\frac{4\pi}{(N+1)^2}} \cdot u(\theta,\az)$},
		xmin=0,xmax=pi,
		ymin=-0.25,ymax=1.05,
		no markers,
		xtick={0, 0.7854, 1.5708, 2.356, 3.1416},
		xticklabels={$0$, $\frac{1}{4}\pi$, $\frac{1}{2}\pi$ ,$\frac{3}{4}\pi$ ,$\pi$},
		legend columns=2,
		legend style={legend cell align=left, /tikz/column 2/.style={column sep=5pt} },
		cycle list name=rwthcolors,
		every axis plot/.append style={thick}
	]

		\addplot table[x=theta,y=N2,col sep=comma] {\figuredatadir/sh_smoothing.csv};
	    \addlegendentry{$N=2$}
		
		\addplot table[x=theta,y=N4,col sep=comma] {\figuredatadir/sh_smoothing.csv};
		\addlegendentry{$N=4$}

		\addplot table[x=theta,y=N8,col sep=comma] {\figuredatadir/sh_smoothing.csv};
		\addlegendentry{$N=8$}
		
		\addplot table[x=theta,y=N16,col sep=comma] {\figuredatadir/sh_smoothing.csv};
		\addlegendentry{$N=16$}
		
		
		\draw[rwthgrau,->] (axis cs:0.75,0.75) -- node[pos=0,anchor=south west]{$N$} (axis cs:0.1,0.1);

	\end{axis}
\end{tikzpicture}%
\tikzaloneEnd%
		\caption{Effect of SH truncation on spatial Dirac delta function}
		\label{fig:sh_smoothing}
	\end{figure}

	For each direction $\vec{s}_{\text{u},q} \in \mathcal{Q}$, we excite $\mathcal{T}$ with
	\begin{align}
		\label{eq:directional_excitation}
		\shcoeffvec{\sigin} = \shcoeffvec{\sigexc} \equiv \sqrt{\frac{4\pi}{(N+1)^2}} \cdot \boldsymbol{y}\herm(\vec{s}_{\text{u},q})
	\end{align}
	which is the band-limited SH equivalent of the spatial Dirac delta function. The normalization is chosen to obtain unit norm $\norm{\shcoeffvec{\sigexc}}_2 = 1$.
	Using the SH addition theorem~\cite{Rafaely2015}, it is easy to show that transforming \eqref{eq:directional_excitation} back to the spatial domain using \eqref{eq:idsht} yields
	\begin{align}
		\sigexc(\inc,\az)
		&= \sum_{n=0}^N \frac{2n+1}{\sqrt{4\pi (N+1)^2}} P_n(\cos \Delta\inc_q),
	\end{align}
	where $P_n(\cdot)$ is the $n$-th degree Legendre polynomial and  $\Delta\inc_q$ is the angle between $(\inc,\az)$ and $(\inc_q,\az_q)$, i.e., $\cos\Delta\inc_q = \cos\inc_q\cos\inc + \cos(\az_q-\az)\sin\inc\sin\inc_q$. We get the maximum value for $\Delta\inc_q=0$. The further trend is shown in Fig.~\ref{fig:sh_smoothing} for various choices of $N$. Obviously, the band-limitation of the Dirac delta impulse results in a spatial blur which decreases with increasing $N$. 
	
	Exciting $\mathcal{T}$ using \eqref{eq:directional_excitation} for each $\vec{s}_{\text{u},q} \in \mathcal{Q}$ yields the responses
	\begin{align}
		\label{eq:directional_response}
		\shcoeffvec{\sigresp}(\vec{s}_{\text{u},q})
		= \sqrt{\frac{4\pi}{(N+1)^2}} \cdot \boldsymbol{T} \boldsymbol{y}\herm(\vec{s}_{\text{u},q}).
	\end{align}
	Using the orthogonality property of matrix $\boldsymbol{Y}_{N,\mathcal{Q}}$ for an adequate sampling grid $\mathcal{Q}$, the matrix $\boldsymbol{T}$ can be reconstructed from the responses $\boldsymbol{\tilde{U}} = \left[\shcoeffvec{\sigresp}(\vec{s}_{\text{u},1}), \ldots, \shcoeffvec{\sigresp}(\vec{s}_{\text{u},Q})\right]$ using
	\begin{align}
		\frac{\sqrt{4\pi(N+1)^2}}{Q} \boldsymbol{\tilde{U}} \boldsymbol{Y}_{N,\mathcal{Q}}
		= \frac{4\pi}{Q} \boldsymbol{T} \boldsymbol{Y}_{N,\mathcal{Q}}\herm \boldsymbol{Y}_{N,\mathcal{Q}}
		= \boldsymbol{T}.
	\end{align}
	Thus, the responses $\boldsymbol{\tilde{U}}$ to the excitations \eqref{eq:directional_excitation} are a complete characterization of $\mathcal{T}$. This will be exploited in the following.
	
	\section[Visualization of Transformations]{Visualization of Transformations\footnote{We provide a MATLAB implementation of the proposed method online at \url{https://www.iks.rwth-aachen.de/qr/visualization-of-sh-domain-operations}}}
	\label{sec:proposed_method}
	
	Our goal is to establish a visual description of the transformation $\mathcal{T}$ which is not only intuitively accessible but also as exhaustive as possible. The spatial domain in $\mathbb{R}^3$ is particularly suitable because of its proximity to human everyday experience. Therefore, we base our method on our findings in Sec.~\ref{sec:linear_operation_characterization}. The idea is to visualize the most important characteristics of the responses from \eqref{eq:directional_response}.
	
	\subsection{Directional Gain}
	
	In classic filter theory, one of the most important properties is the magnitude response. Analogously, we define a directional gain which describes the sensitivity of the filter to an excitation from a specific direction,
	\begin{align}
		\eta(\inc,\az) \equiv \eta(\uvec{s}) = \frac{\norm{\shcoeffvec{\sigresp}(\uvec{s})}_2}{\norm{\shcoeffvec{\sigexc}(\uvec{s})}_2}= \norm{\shcoeffvec{\sigresp}(\uvec{s})}_2.
	\end{align}
	A value of $\eta(\uvec{s})=1$ means that the energy fed into the system from the corresponding direction $\uvec{s}$ is preserved. Lower values indicate an attenuation, larger values an amplification.
	
	\subsection{Response Energy Vector}
	
	While $\eta(\uvec{s})$ is an appropriate measure for spatially selective filters, it fails to encode the spatial redistribution of energy. We tackle this aspect by measuring the energy centroid of the response for an excitation from direction $\uvec{s}$,~\cite{Daniel2001}
	\begin{align}
		\label{eq:energy_vector}
		\vec{r}_\text{E}(\uvec{s}) = \frac{ \int_\unitsphere \abs{\boldsymbol{y}_{\tilde{N}}(\puvec{s}')\shcoeffvec{\sigresp}(\uvec{s})}^2 \puvec{s}'\,\diff{}\puvec{s}' }{ \int_\unitsphere \abs{\boldsymbol{y}_{\tilde{N}}(\puvec{s}')\shcoeffvec{\sigresp}(\uvec{s})}^2 \,\diff{}\puvec{s}' }.
	\end{align}
	The enumerator determines the \enquote{center of mass} of the response energy which is further normalized using the expression in the denominator.

	Note that in spatial audio signal processing, this energy vector was originally proposed as a very simple summing localization model to predict the perceived direction of sound in a multi-loudspeaker scenario~\cite{Gerzon1992,Daniel2001}.
	
	Both the direction and norm of $\vec{r}_\text{E}(\uvec{s})$ play an important role to describe $\mathcal{T}$. If $\abs{\vec{r}_\text{E}(\uvec{s})}$ is about one, the response is very focused in the direction of $\vec{r}_\text{E}(\uvec{s})$. The lower the value of $\abs{\vec{r}_\text{E}(\uvec{s})}$ is, the more spatially blurred is the response. For low norms, the direction of $\vec{r}_\text{E}(\uvec{s})$ looses its  information content.
	For instance, this is the case for the operation
	\begin{align}
		\label{eq:problematic_operation}
		\boldsymbol{T} = \int_\unitsphere \left(\boldsymbol{y}_{N}\herm(\uvec{s})+\boldsymbol{y}_{N}\herm(-\uvec{s})
		\right)\boldsymbol{y}_{N}(\uvec{s})
		\,\diff{}\uvec{s},
	\end{align}
	which is the superposition of an identity and a mirror operation. As a consequence, the response to a directed excitation is highly directed in two opposing directions, but $\abs{\vec{r}_\text{E}(\uvec{s})}=0\,\forall \uvec{s}$. Therefore, the direction $\vec{r}_{\text{E},\text{u}}(\uvec{s})$ is not defined.
	Although the response energy vector does not provide information for special cases like Eq.~\eqref{eq:problematic_operation}, it is useful for most practical use-cases as will be shown in Sec.~\ref{sec:evaluation}. We refrain from using a more sophisticated measure for the distribution of the spatial response in favor of simplicity and readability of our proposed visualization method.

	At a first glance, the norm $\abs{\vec{r}_\text{E}(\uvec{s})}$ seems to be limited by 1. However, the actual maximum value is lower for finite $\tilde{N}$.
	The values of $\abs{\vec{r}_\text{E}(\uvec{s})}$ lie in the range $ 0 \le \abs{\vec{r}_\text{E}(\uvec{s})} \le r_{\text{E,max},\tilde{N}}$ with~\cite{Daniel2001}
	\begin{align}
		r_{\text{E,max},\tilde{N}} = \max \{ x | P_{\tilde{N}+1}(x) = 0\}.
	\end{align}
	
	Although an excitation of the form of \eqref{eq:directional_excitation} yields \textit{maximum directivity} in the sense that it has the largest value in look direction among all possible excitation signals with identical energy, it exhibits a lower value of $\abs{\vec{r}_\text{E}(\uvec{s})}$ then $r_{\text{E,max},\tilde{N}}$.
	Assuming $\tilde{N} = N$, an identity transformation $\boldsymbol{T} = \boldsymbol{I}$ yields~\cite{Daniel2001}
	\begin{align}
		\label{eq:energy_vector_truncated_dirac}
		\abs{\vec{r}_\text{E}(\uvec{s})} = \frac{N}{N+1} \quad\forall\uvec{s}.
	\end{align}
	Values for different $N$ are listed in Table~\ref{tbl:max_re_normalization}.
	
	\begin{table}
		\caption{Values for $\frac{N}{N+1}$ and $r_{\text{E,max},N}$}
		\label{tbl:max_re_normalization}
		\centering
		\begin{tabular}{ccc}\toprule
			$N$ & $\frac{N}{N+1}$& $r_{\text{E,max},N}$ \\\midrule
			1 & 0.500 & 0.577 \\
			2 & 0.667 & 0.775 \\
			3 & 0.750 & 0.861 \\
			4 & 0.800 & 0.906 \\
			5 & 0.833 & 0.932 \\
			6 & 0.857 & 0.949 \\
			7 & 0.875 & 0.960 \\
			8 & 0.889 & 0.968 \\
			9 & 0.900 & 0.974 \\
			10 & 0.909 & 0.978 \\
			\bottomrule
		\end{tabular}
	\end{table}

	\begin{figure*}[t!]
		\centering
		\setlength{\abovecaptionskip}{0.5ex}
		\setlength\figurewidth{.76\linewidth}
		\setlength\figureheight{.32\linewidth}
		\newcommand{\simulationname}{transformation_rot}
		\input{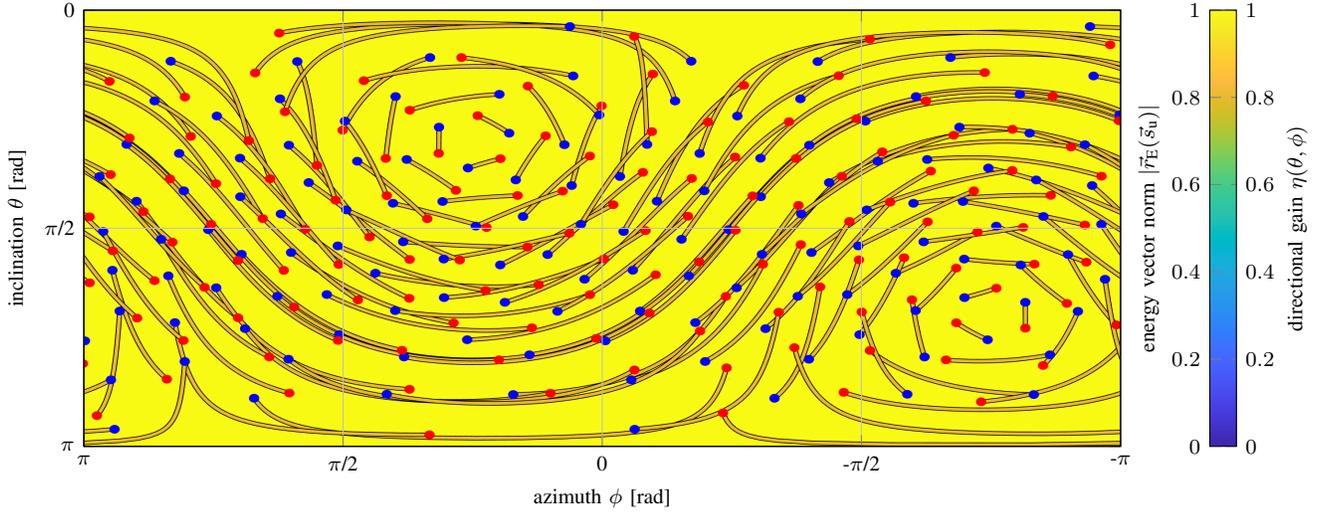}
		\caption{Visualization of rotation operation of $+60^\circ$ around the axis $\vec{r}=\frac{1}{\sqrt{3}}(1,1,1)\trans$}
		\label{fig:transformation_rot}
	\end{figure*}
	
	\subsection{Visualization}
	
	$\eta(\uvec{s})$ is a scalar function defined on the unit sphere, which can be visualized using standard tools. The unit sphere can be flattened for print or screen display using an appropriate map projection~\cite{Bernschuetz2012} such as an equirectangular one. In an equirectangular projection, azimuth $\az$ and inclination $\inc$ are mapped to the axes of a standard Cartesian coordinate system which is simple and also accessible to a non-specialist reader due to the general familiarity of world maps. It is convenient to encode $\eta(\uvec{s})$ using an appropriate color map, preferably one optimized for a uniform perceptual contrast to prevent presuming false anomalies in the data~\cite{Kovesi2015}.
	
	When it comes to the energy vector $\vec{r}_\text{E}(\uvec{s})$, a vector field has to be visualized. We propose to evaluate $\vec{r}_\text{E}(\uvec{s})$ on a uniformly sampled coarse grid. In the following, we use a grid with $Q=144$ sampling points taken from \cite{HardinSloane1996}.
	The vector field is visualized on top of the axis showing $\eta(\uvec{s})$. Each $\vec{s}_{\text{u},q}$ is shown as a \textcolor{blue}{blue} mark, the corresponding $\vec{r}_{\text{E},\text{u}}(\vec{s}_{\text{u},q})$ as a \textcolor{red}{red} mark (\crazylinelegend).
	The norm $\abs{\vec{r}_\text{E}(\inc_q,\az_q)}$ is encoded in the color of a line which connects both marks. This line is drawn along the shortest path between $\vec{s}_{\text{u},q}$ and $\vec{r}_{\text{E},\text{u}}(\vec{s}_{\text{u},q})$ on $\unitsphere$ which is the \textit{great circle} segment
	\begin{align}
		\label{eq:great_circle_trajectory}
		\vec{r}_q(t) = \frac{\sin\left( (1-t)\Delta\right)}{\sin\Delta} \vec{r}_{\text{E},\text{u}}(\vec{s}_{\text{u},q}) + \frac{\sin\left( t\Delta \right)}{\sin\Delta} \vec{s}_{\text{u},q}
	\end{align}
	parametrized with $0 \le t \le 1$ and $\Delta = \arccos\langle \vec{r}_{\text{E},\text{u}}(\vec{s}_{\text{u},q}) , \vec{s}_{\text{u},q} \rangle$. Here, $\langle\cdot,\cdot\rangle$ denotes the inner vector product.

	The following section clarifies the proposed visualization method by giving several examples.

	\section{Evaluation}
	\label{sec:evaluation}

	We base our evaluation on different transformation techniques of practical relevance which were proposed in the past. These examples are chosen in a way that different aspects of the proposed visualization method are in focus. 
	
	\subsection{Spherical Harmonics Rotation}

	\paragraph*{Functional description}
	
	In the following, we consider coefficients $\shcoeff{\sigin}$ given in a Cartesian reference frame with basis vectors $(\vec{e}_\mathrm{x},\vec{e}_\mathrm{y},\vec{e}_\mathrm{z})$ which shall be described in a rotated reference frame with basis vectors $(\vec{e}_{\tilde{\mathrm{x}}},\vec{e}_{\tilde{\mathrm{y}}},\vec{e}_{\tilde{\mathrm{z}}})$. Both reference frames share the same origin.
	
	We use a rotation matrix $\boldsymbol{\mathcal{R}}$, i.e., a matrix which fulfills $ \langle \boldsymbol{\mathcal{R}}\cdot\vec{r}_1 , \boldsymbol{\mathcal{R}}\cdot\vec{r}_2 \rangle = \langle \vec{r}_1 , \vec{r}_2 \rangle\ \forall \vec{r}_1,\vec{r}_2\in\mathbb{R}^3$ and $\det \boldsymbol{\mathcal{R}} = 1$, to describe the mapping $\vec{r} \mapsto \vec{\tilde{r}}$ such that
	\begin{align}
		\vec{\tilde{r}} = \boldsymbol{\mathcal{R}} \cdot \vec{r}.
	\end{align}
	This transformation exhibits three degrees of freedom which are often expressed in terms of Euler angles yaw, pitch, roll or quaternions for convenience.
	
	The SH signal in the rotated reference frame is then given by the linear relation~\cite{GumerovDuraiswami2004}
	\begin{align}
		\label{eq:sh_rot}
		\tilde{x}_{nm} = \sum_{m'=-n}^n R_n^{mm'}(\boldsymbol{\mathcal{R}})  x_{nm'}
	\end{align}
	The rotation coefficients $R_n^{mm'}(\boldsymbol{\mathcal{R}})$ can be obtained via their integral representation~\cite{GumerovDuraiswami2004}
	\begin{align}
		\label{eq:sh_rot_integral_representation}
		R_n^{mm'}(\boldsymbol{\mathcal{R}}) = \int_\unitsphere Y_n^{m'}(\boldsymbol{\mathcal{R}}^{-1}\cdot\uvec{\tilde{s}}) \left\lbrack Y_n^m(\uvec{\tilde{s}})\right\rbrack^* \,\diff{}\uvec{\tilde{s}}.
	\end{align}
	with $\uvec{\tilde{s}} = \boldsymbol{\mathcal{R}} \cdot \uvec{s}$. Here, $(\cdot)\inv$ denotes the matrix inverse. Note that in practice, it is often more convenient to obtain the $R_n^{mm'}(\boldsymbol{\mathcal{R}})$ using recurrence relations which is less intuitive but computationally more efficient than evaluating \eqref{eq:sh_rot_integral_representation}~\cite{IvanicRuedenberg1996}.

	For an order-limited coefficient vector $\shcoeffvec{\sigin}$, \eqref{eq:sh_rot} and \eqref{eq:sh_rot_integral_representation} can be rewritten in matrix notation in the manner of \eqref{eq:sh_transformation} using
	\begin{align}
		\label{eq:sh_rot_matrix}
		\boldsymbol{T}^\text{(rot)}
		&= \int_\unitsphere \boldsymbol{y}_{\tilde{N}}\herm(\uvec{\tilde{s}}) \boldsymbol{y}_{N}(\boldsymbol{\mathcal{R}}^{-1}\cdot\uvec{\tilde{s}}) \,\diff{}\uvec{\tilde{s}}.
	\end{align}

	\paragraph*{Visualization}
	
	Fig.~\ref{fig:transformation_rot} shows $\boldsymbol{T}^\text{(rot)}$ for a rotation of $+\frac{\pi}{3} \,\hat{=}\, +60^\circ$ around the axis $\vec{r}=\frac{1}{\sqrt{3}}(1,1,1)\trans$ for an input and output truncation order of $\tilde{N}=N=4$. The choice of $N=4$ is a typical choice in the context of microphone-array-recorded spatial audio processing (e.g.,~\cite{KentgensKuehlAntweilerEtAl2018}).
	
	The first thing to notice is the uniformly colored background indicating $\boldsymbol{T}^\text{(rot)}$'s spatial all-pass characteristic, i.e., $\eta(\inc,\az)=1 \,\forall\inc,\az$. The operation is therefore rotation invariant for input signals with respect to the energy at the output.
	
	The trajectories (\crazylinelegend) connecting $\uvec{s}$ and $\vec{r}_{\text{E},\text{u}}(\uvec{s})$ illustrate the rotation well. The rotation axis, i.e., $\vec{r}_{\text{E},\text{u}}(\uvec{s})=\uvec{s}$, is easy to see as the red and blue marks nearby are close to each other. This is due to the fact that these directions exhibit a less severe positional change in the sense of the great circle distance compared to other directions.
	Furthermore, it is obvious now that the great-circle-distance-based trajectory definition in \eqref{eq:great_circle_trajectory} results in an easier-to-interpret visualization than the shortest connection in the $(\inc,\az)$ plane.
	
	The identical color of all trajectory lines indicate that $\abs{\vec{r}_\text{E}(\uvec{s})} = \frac{N}{N+1} = 0.8$. This means that the impulse shape of a directional excitation is maintained at the output (cf.~\eqref{eq:energy_vector_truncated_dirac}).

	\begin{figure*}[t!]
		\centering
		\setlength{\abovecaptionskip}{0.5ex}
		\setlength\figurewidth{.76\linewidth}
		\setlength\figureheight{.32\linewidth}
		\newcommand{\simulationname}{transformation_warp}
		\input{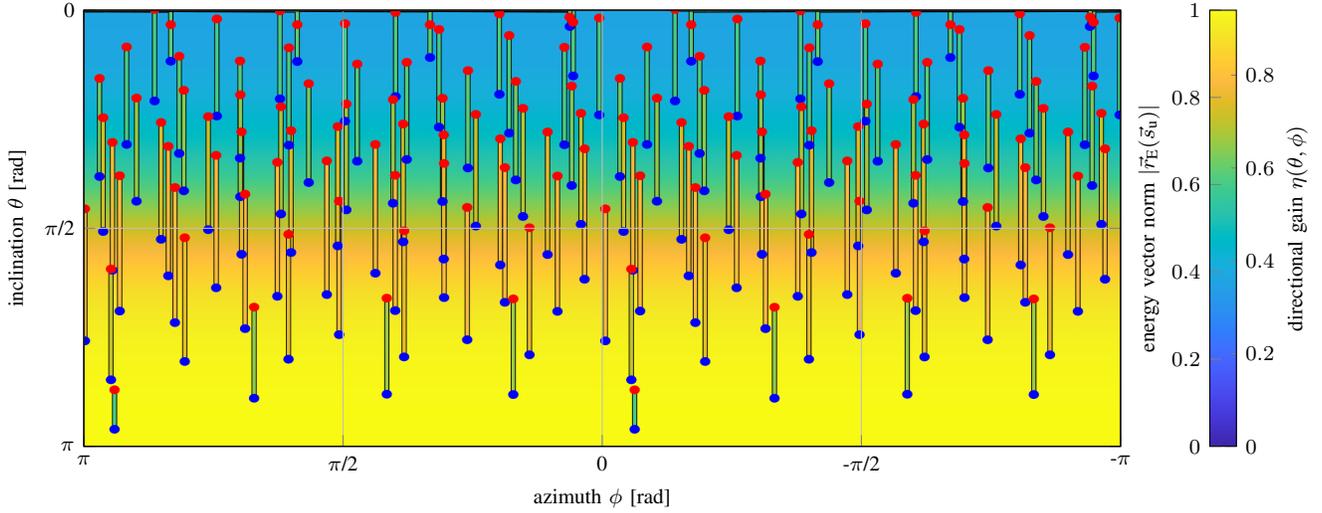}
		\caption{Visualization of space warping operation}
		\label{fig:transformation_warp}
	\end{figure*}
	
	\subsection{Space Warping}

	\paragraph*{Functional description}

	\textit{Space warping} is a sound field modification technique which is originally accredited to Pomberger and Zotter~\cite{PombergerZotter2011} as a generalization of Gerzon's forward dominance operation~\cite{GerzonBarton1992}. The underlying idea is to squeeze or stretch parts of the axes of the spherical coordinate system in order to modify the spatial characteristics of the given signal.
	
	One variant is space warping along the inclination angle $\inc$ which is defined in the spatial domain as~\cite{KentgensJax2020}
	
	\begin{align}
		\label{eq:space_warping_along_inclination}
		\boldsymbol{T}^\text{(warp)}
		&= \int_\unitsphere g(\tilde{\inc}) \boldsymbol{y}_{\tilde{N}}\herm(\tilde{\inc},\az)
		\boldsymbol{y}_{N}(f^{-1}(\tilde{\inc}),\az)
		\,\sin\tilde{\inc}\diff\tilde{\inc}\diff\az.
	\end{align}
	Here, $f^{-1}$ denotes the inverse of the monotone warping function $f{:\ } (0,\pi)\rightarrow(0,\pi)$. We follow \cite{PombergerZotter2011} and choose
	\begin{align}
		\label{eq:warping_function}
		f(\inc)
		&= \cos^{-1}\left( \frac{\cos(\inc)+\alpha}{1+\alpha\cos(\inc)} \right).
	\end{align}
	Parameter $\alpha$ with $-1 < \alpha < 1$ controls the amount of warping. Choosing the equalization function $g(\tilde{\inc})$ as
	\begin{align}
		\label{eq:equalization_function}
		g(\tilde{\inc})
		&= \frac{\sqrt{1-\alpha^2}}{1-\alpha\cos(\tilde{\inc})} %
	\end{align}
	yields preservation of the energy of the warped signal.
	
	Eqs.~\eqref{eq:sh_rot_matrix} and \eqref{eq:space_warping_along_inclination} share a similar structure. Both operations apply an angular modification on the variable of integration. However, the angular modification in \eqref{eq:space_warping_along_inclination} is non-uniform and an additional gain $g(\tilde{\inc})$ is applied. Therefore, we expect several differences in the visualization.

	\paragraph*{Visualization} Fig.~\ref{fig:transformation_warp} shows $\boldsymbol{T}^\text{(warp)}$ for a truncation order of $\tilde{N}=N=4$ and $\alpha=0.8$. As expected, $\vec{r}_{\text{E},\text{u}}(\uvec{s})$ exhibits a modification with respect to $\uvec{s}$ along $\inc$ only. The length of the trajectory lines shows that the angular modification is stronger at the equator $\inc=\pi/2$ than at the \enquote{north} and \enquote{south poles} at $\inc=0$ and $\inc=\pi$, respectively. In order to preserve the energy at the output, the gain function \eqref{eq:equalization_function} counteracts the spatially concentrating behavior of the warping as can be seen from the visualization of $\eta(\uvec{s})$. Those excitations from $\uvec{s}$ for which the red marks show a high spatial density are attenuated while those excitations from $\uvec{s}$ which yield a low density of red marks are amplified.
	
	Moreover, we can observe that not all trajectory lines are yellow anymore, i.e., $\abs{\vec{r}_\text{E}(\uvec{s})}<\frac{N}{N+1}$. This indicates spatial blurring of the response and is due to the fact that the warping operation is not order preserving and only a finite truncation order $N$ is considered.

	\begin{figure*}[t!]
		\centering
		\setlength{\abovecaptionskip}{0.5ex}
		\setlength\figurewidth{.76\linewidth}
		\setlength\figureheight{.32\linewidth}
		\newcommand{\simulationname}{transformation_nr_dp}
		\input{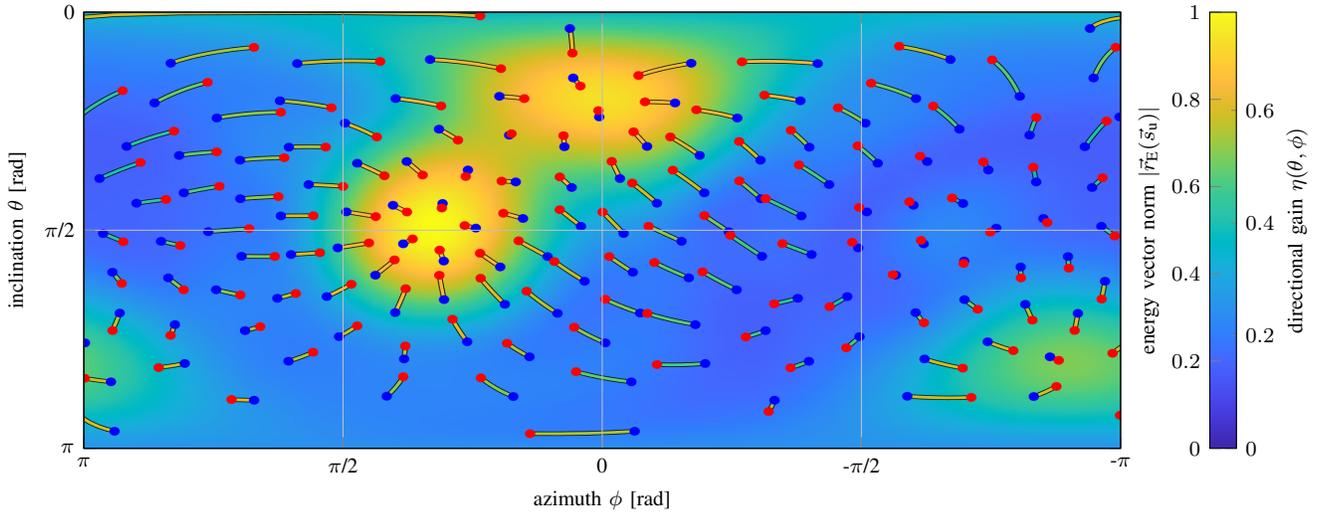}
		\caption{Visualization of direction-preserving noise reduction}
		\label{fig:transformation_nr_dp}
	\end{figure*}
	
	\begin{figure*}[t!]
		\centering
		\setlength{\abovecaptionskip}{0.5ex}
		\setlength\figurewidth{.76\linewidth}
		\setlength\figureheight{.32\linewidth}
		\newcommand{\simulationname}{transformation_nr_mbf}
		\input{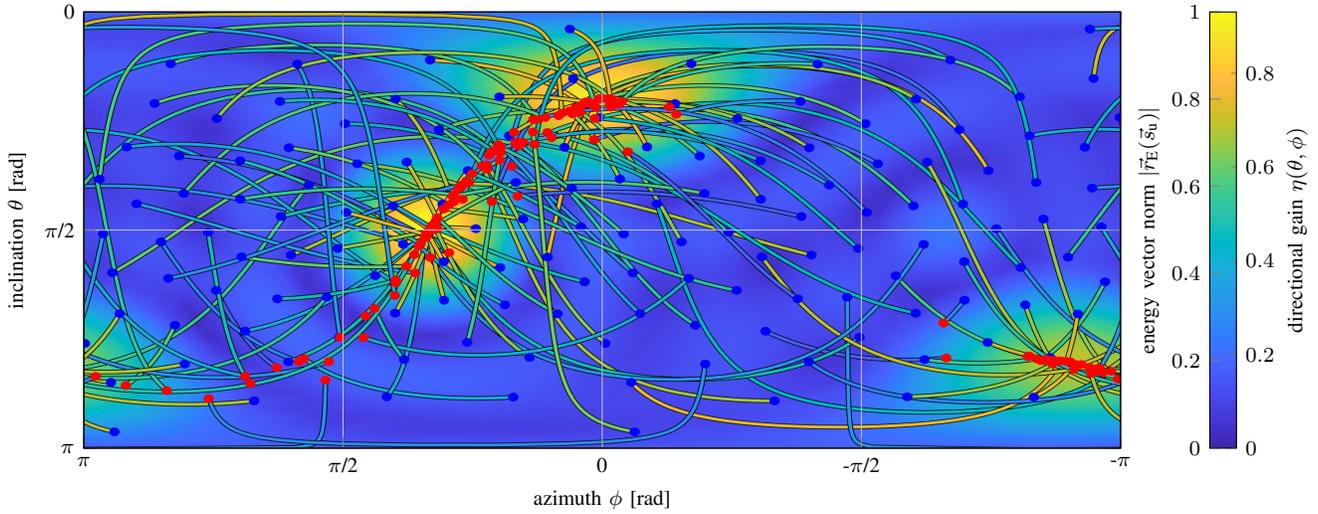}
		\caption{Visualization of multi-beamformer noise reduction}
		\label{fig:transformation_nr_mbf}
	\end{figure*}
	
	\subsection{Wiener Filter}
	
	\paragraph*{Functional description}
	
	Given a signal model
	\begin{align}
		\shcoeffvec{\sigin} = \shcoeffvec{d} + \shcoeffvec{\nu}
	\end{align}
	with desired random signal $\shcoeffvec{d}$ and uncorrelated additive noise $\shcoeffvec{\nu}$, we would like to apply a noise reduction filter in the form of \eqref{eq:sh_transformation} to obtain an estimate of the desired signal such that $\shcoeffvec{\sigout}\approx\shcoeffvec{d}$. We assume that the signals themselves are unknown but their second-order statistical properties are known in form of their covariance matrices $\boldsymbol{\Phi}_\mathrm{d} = \Expect\{\shcoeffvec{d}\shcoeffvec{d}\herm\}$ and $\boldsymbol{\Phi}_\mathrm{\nu} = \Expect\{\shcoeffvec{\nu}\shcoeffvec{\nu}\herm\}$. Here, $\Expect\{\cdot\}$ denotes mathematical expectation.
	
	In the following, we consider two variants of a Wiener filter for this purpose which aim at preserving the spatial characteristics of the desired signal. The \textit{direction preserving} approach proposed by Herzog and Habets in \cite{HerzogHabets2019} applies gains $h(\inc,\az)$ in the spatial domain. Its integral representation
	\begin{align}
		\boldsymbol{T}^\text{(NR-DP)} = \int_\unitsphere h(\inc,\az) \boldsymbol{y}_{N}\herm(\inc,\az) \boldsymbol{y}_{N}(\inc,\az)\,\sin\inc\diff\inc\diff\az
	\end{align}
	has a structure which resembles \eqref{eq:space_warping_along_inclination} but does not apply any angular warping but a gain function only.

	The spatial gains have the typical structure of a Wiener filter,
	\begin{align}
		\label{eq:nr_spatial_gain}
		h(\inc,\az) = \frac{\Phi_\mathrm{d}(\inc,\az)}{\Phi_\mathrm{d}(\inc,\az)+\mu\Phi_\mathrm{\nu}(\inc,\az)}.
	\end{align}
	The $\Phi_{(\cdot)}(\inc,\az)$ are the desired \textit{steered-response powers} of the signal and noise, respectively,
	\begin{align}
		\Phi_{(\cdot)}(\inc,\az) = \boldsymbol{y}_{N}(\inc,\az) \boldsymbol{\Phi}_{(\cdot)} \boldsymbol{y}_{N}\herm(\inc,\az).
	\end{align}
	Parameter $\mu$ allows to control the trade-off between distortion of the desired signal and aggressiveness of the noise reduction. In the following, we choose $\mu=1$.

	An even simpler form of a Wiener filter is the \textit{matrix parametric multichannel Wiener filter} (e.g.,~\cite{DocloMoonen2002,HerzogHabets2019,Pratt1972})
	\begin{align}
		\label{eq:nr_mbf}
		\boldsymbol{T}^\text{(NR-PM)} = \boldsymbol{\Phi}_\mathrm{d} \left( \boldsymbol{\Phi}_\mathrm{d} + \mu \boldsymbol{\Phi}_\mathrm{\nu} \right)\inv
	\end{align}
	which again resembles \eqref{eq:nr_spatial_gain} but is a vectorial operation directly defined in the SH domain. Given that $\shcoeffvec{d}$ is a linear combination of $J<(N+1)^2$ uncorrelated sources, $\boldsymbol{T}^\text{(NR-PM)}$ can be interpreted as an \textit{beamform-and-project} approach where the source signals are first extracted by a projection on a $J$-dimensional subspace which is again projected back on the signal subspace.
	
	\begin{table}
		\centering
		\caption{Source properties}
		\label{tbl:nr_source_properties}
		\newcolumntype{L}[1]{>{\raggedright\let\newline\\\arraybackslash\hspace{0pt}}m{#1}}
		\newcolumntype{C}[1]{>{\centering\let\newline\\\arraybackslash\hspace{0pt}}m{#1}}
		\newcolumntype{R}[1]{>{\raggedleft\let\newline\\\arraybackslash\hspace{0pt}}m{#1}}
		\renewcommand{\arraystretch}{1.3}
		\setlength{\tabcolsep}{0ex}
		\begin{tabular}{C{6ex}C{10ex}*{2}{R{5ex}@{}L{5ex}}}
			\toprule
			$i$ & $a_i$ & \multicolumn{2}{c}{$\inc_i$} & \multicolumn{2}{c}{$\az_i$} \\ 
			\midrule
			1 & 0.8 & $\frac{1}{5}$&$\pi$ & \multicolumn{2}{c}{$0$} \\
			2 & 1   & $\frac{1}{2}$&$\pi$ & $\frac{1}{3}$&$\pi$ \\
			3 & 0.4 & $\frac{4}{5}$&$\pi$ & $-\frac{9}{10}$&$\pi$ \\ 
			\bottomrule
		\end{tabular}
	\end{table}
	
	We now consider three uncorrelated sources with amplitudes and directions as given in Table~\ref{tbl:nr_source_properties}. The desired signal covariance matrix then is
	\begin{align}
		\boldsymbol{\Phi}_\mathrm{d} = \sum_{i=1}^3\abs{a_i}^2\boldsymbol{y}_{N}\herm(\inc_i,\az_i) \boldsymbol{y}_{N}(\inc_i,\az_i).
	\end{align}
	Moreover, we consider the noise signal $\shcoeffvec{\nu}$ to be diffuse, i.e., $\boldsymbol{\Phi}_\mathrm{\nu} \propto \boldsymbol{I}$. The signal-to-noise ratio is chosen to be $10 \lg \frac{\trace\boldsymbol{\Phi}_\mathrm{d}}{\trace\boldsymbol{\Phi}_\mathrm{\nu}} \,\mathrm{dB} = 0\,\mathrm{dB}$.

	\paragraph*{Visualization} The filters $\boldsymbol{T}^\text{(NR-DP)}$ and $\boldsymbol{T}^\text{(NR-PM)}$ for an input and output order of $\tilde{N}=N=4$ are visualized in Figs.~\ref{fig:transformation_nr_dp} and \ref{fig:transformation_nr_mbf}, respectively. The first thing to notice in both plots is that $\eta(\uvec{s})$ is high (yellow) only in the regions around the source directions and low (blue) otherwise, which is the expected behavior for the noise reduction. However, there are notable differences between the methods in detail.

	In Fig.~\ref{fig:transformation_nr_dp}, the blue and red dots are relatively close to each other both for the desired and the unwanted directions. Excitations from the unwanted directions are almost equally attenuated. Thus, using $\boldsymbol{T}^\text{(NR-DP)}$, the residual noise preserves its spatial distribution.
	
	In contrast, the residual noise in Fig.~\ref{fig:transformation_nr_mbf} shows different characteristics. The noise attenuation is stronger but significantly less isotropical. In addition, the residual noise is focused in the directions of the desired sources which can be seen from the distribution of the red marks. This can be well explained using the beamform-and-project interpretation of $\boldsymbol{T}^\text{(NR-PM)}$. The input signal $\boldsymbol{x_{nm}}$ is projected onto a three-dimensional subspace. In this domain, the desired signal and residual noise are mixed and no spatial information on the noise characteristics is maintained. When projecting the intermediate signal back to the original signal subspace, the residual noise shows the same spatial characteristics as the desired signal.
	
	In total, our interpretation of the characteristics of $\boldsymbol{T}^\text{(NR-DP)}$ and $\boldsymbol{T}^\text{(NR-PM)}$ is well in line with the experimental findings in \cite{HerzogHabets2019}.

	\section{Discussion}
	\label{sec:discussion}
	
	The examples in Sec.~\ref{sec:evaluation} reveal the potential of the proposed visualization method. For many practical applications, the method allows for deep insights in the behavior of a spatial operation. However, there are cases, in which the visualization might require modifications to convey all aspects of $\mathcal{T}$.
	
	The choice of $Q=144$ sampling points for the visualization of $\vec{r}_{\text{E},\text{u}}(\uvec{s})$ in Sec.~\ref{sec:evaluation} was empirically determined as a good compromise of spatial resolution for the given transformations of input and output order $N=4$ and readability in a two-column graphic in an A4- or letter-sized paper. Naturally, the grid can be adapted to the specific needs of the user. Hereby, it is important to choose a sufficiently high $Q$ in order to not violate the sampling theorem as discussed in Sec.~\ref{sec:linear_operation_characterization}. Also, we recommend to choose a (quasi-)uniformly sampled grid which does not give visual preference to specific directions and also provides an intuitive way to understand the distortion of proportions in the map projection.

	Moreover, to clean up the visualization and to prevent misinterpretation of $\vec{r}_{\text{E},\text{u}}(\uvec{s})$, an improvement could be to discard those red marks and their respective trajectory lines for which $\abs{\vec{r}_\text{E}(\uvec{s})}$ is below a certain threshold. An alternative is to refrain from color-coding $\abs{\vec{r}_\text{E}(\uvec{s})}$ but to use different line widths instead. 
	
	Furthermore,  several other modifications of the visualization method can be applied dependent on the operation under investigation and the application. The following should be mentioned at this point: for example, a logarithmic visualization of $\eta(\inc,\az)$ (in dB), other map projection methods~\cite{Maling1992,Bernschuetz2012}, or a normalization of $\abs{\vec{r}_\text{E}(\uvec{s})}$ by $\frac{N}{N+1}$ to yield unit values for $\boldsymbol{T}=\boldsymbol{I}$.
	
	Dependent on the application, it might be necessary to visualize a transformation $\mathcal{T}$ over a certain dimension such as time or frequency. Although this is not directly possible with the proposed method, this could be done with an animated or interactive plot.

	\section{Conclusion}

	In the present paper, we proposed a method for the concise, unified visualization of linear operations and systems in the SH domain. We identified different key properties of such operations which our approach is based on. In an evaluation with different practical examples, our method has proven to be a useful tool for the interpretation of the behavior of the systems. We discussed the further potential and limitations of our proposition.
	By promoting a novel tool for the analysis and comparison of SH-domain systems and operations, we hope to contribute to the understanding of such systems and to build a foundation for future discussions.
	
	\bibliographystyle{IEEEtran}
	\bibliography{literature.bib}

\end{document}